%% file: HPS_outline.tex
\documentclass[12pt,tightenlines,superscriptaddress]{revtex4-2}

\usepackage[page,title]{appendix}
\usepackage{url}
\usepackage{graphicx}
\usepackage{amssymb}
\usepackage{bm}
\usepackage{mathrsfs}
\usepackage{amsmath}
\usepackage{amsfonts}
\usepackage{color}
\usepackage{xspace}
\usepackage{tikz}
\usepackage{slashed}
\usepackage{cancel}
\usepackage{hyperref}
\usepackage[utf8]{inputenc}											
\usepackage{slashed}

\usetikzlibrary{arrows}
\usetikzlibrary{shapes}
\usetikzlibrary{trees}
\usetikzlibrary{matrix}
\usetikzlibrary{arrows}

\usepackage{graphicx,epsfig,psfrag,bm,amssymb}
\usepackage{dcolumn}
\usepackage{bm}
\usepackage{color}
\usepackage{mathrsfs,amsfonts,hepunits,color}
\usepackage[utf8]{inputenc}
\usepackage{epsfig,latexsym,cancel,amssymb,amsmath,mathrsfs}
\usepackage{graphicx}
\usepackage{epstopdf}
\usepackage{mciteplus}
\usepackage{latexsym}
\usepackage{amsthm}
\usepackage{amsmath}
\usepackage{amssymb}
\usepackage{hepunits}
\usepackage{hyperref}
\usepackage{bbm}
\usepackage{bm}
\usepackage{xfrac}
\usepackage{colordvi}
\usepackage{comment}
\usepackage{dcolumn}
\usepackage{times,latexsym,graphicx,wrapfig}
\usepackage{epsfig,lineno,bm}
\usepackage{slashed}
\usepackage{booktabs}
\usepackage{relsize}

\usepackage{siunitx}
\usepackage{mhchem}
  
\newcommand{\be}{\begin{eqnarray}}
\newcommand{\ee}{\end{eqnarray}}

\newcommand{\benum}{\begin{enumerate}}
\newcommand{\eenum}{\end{enumerate}}
\newcommand{\bi}{\begin{itemize}}
\newcommand{\ei}{\end{itemize}}

\newcommand{\aprime}{$A^{\prime}$\xspace}
\def\babar{\mbox{\slshape B\kern-0.1em{\smaller A}\kern-0.1emB\kern-0.1em{\smaller A\kern-0.2em R}}}

\colorlet{RED}{red}
\colorlet{BLUE}{blue}
\colorlet{ORANGE}{orange}

\newcommand\snowmass{\begin{center}\rule[-0.2in]{\hsize}{0.01in}\\\rule{\hsize}{0.01in}\\
\vskip 0.1in Submitted to the  Proceedings of the US Community Study\\ 
on the Future of Particle Physics (Snowmass 2021)\\ 
\rule{\hsize}{0.01in}\\\rule[+0.2in]{\hsize}{0.01in} \end{center}}

\newcommand{\maprime}{A^{\prime}}

\newcommand{\epem}{$\mathrm{e}^{+}\mathrm{e}^{-}$\xspace}
\newcommand{\eminus}{$\mathrm{e}^{-}$\xspace}

\DeclareUnicodeCharacter{2212}{-} 

\begin{document}


%
%
\title{The Heavy Photon Search Experiment}

\date{\today}
\hspace*{0pt}\hfill 

\bigskip

\input{authorabstract}

\maketitle

\snowmass{}

\newpage
\tableofcontents
\newpage


\clearpage
\section{Introduction}
\input{introduction}
\section{Experimental Overview}
\input{experimental_overview}
 \clearpage
 \section{Status and Plans}
 \input{status_plans}
\bibliography{bibliography}

\end{document}

%% file: authorabstract.tex
\author{Nathan~Baltzell}
\affiliation{Thomas Jefferson National Accelerator Facility, Newport News, Virginia 23606}

\author{Marco~Battaglieri}
\affiliation{Thomas Jefferson National Accelerator Facility, Newport News, Virginia 23606}

\author{Mariangela~Bondi}
\affiliation{Istituto Nazionale di Fisica Nucleare, Sezione di Genova e Dipartimento di Fisica dell'Universit\`a, 16146 Genova, Italy}

\author{Sergei~Boyarinov}
\affiliation{Thomas Jefferson National Accelerator Facility, Newport News, Virginia 23606}

\author{Cameron~Bravo}
\affiliation{SLAC National Accelerator Laboratory, Menlo Park, CA 94025, USA}

\author{Stephen~Bueltmann}
\affiliation{Old Dominion University, Norfolk, Virginia 23529}

\author{Volker~Burkert}
\affiliation{Thomas Jefferson National Accelerator Facility, Newport News, Virginia 23606}

\author{Pierfrancesco~Butti}
\affiliation{SLAC National Accelerator Laboratory, Menlo Park, CA 94025, USA}

\author{Tongtong~Cao}
\affiliation{University of New Hampshire, Department of Physics, Durham, NH 03824}

\author{Massimo~Carpinelli}
\affiliation{Universit\`a di Sassari e Istituto Nazionale di Fisica Nucleare, 07100 Sassari, Italy}

\author{Andrea~Celentano}
\affiliation{Istituto Nazionale di Fisica Nucleare, Sezione di Genova e Dipartimento di Fisica dell'Universit\`a, 16146 Genova, Italy}

\author{Gabriel~Charles}
\affiliation{Institut de Physique Nucleaire d'Orsay, IN2P3, BP 1, 91406 Orsay, France}

\author{Chris~Cuevas}
\affiliation{Thomas Jefferson National Accelerator Facility, Newport News, Virginia 23606}

\author{Annalisa~D'Angelo}
\affiliation{Istituto Nazionale di Fisica Nucleare, Sezione di Roma-TorVergata e Dipartimento di Fisica dell'Universit\`a, Roma, Italy}

\author{Domenico~D’Urso}
\affiliation{Universit\`a di Sassari e Istituto Nazionale di Fisica Nucleare, 07100 Sassari, Italy}

\author{Natalia~Dashyan}
\affiliation{Yerevan Physics Institute, 375036 Yerevan, Armenia}

\author{Marzio~De Napoli}
\affiliation{Istituto Nazionale di Fisica Nucleare, Sezione di Catania e Dipartimento di Fisica dell'Universit\`a, Catania, Italy}

\author{Raffaella~De~Vita}
\affiliation{Istituto Nazionale di Fisica Nucleare, Sezione di Genova e Dipartimento di Fisica dell'Universit\`a, 16146 Genova, Italy}

\author{Alexandre~Deur}
\affiliation{Thomas Jefferson National Accelerator Facility, Newport News, Virginia 23606}

\author{Miriam~Diamond}
\affiliation{University of Toronto, Toronto, ON M5S, Ca}

\author{Raphael~Dupre}
\affiliation{Institut de Physique Nucleaire d'Orsay, IN2P3, BP 1, 91406 Orsay, France}

\author{Rouven~Essig}
\affiliation{Stony Brook University, Stony Brook, NY 11794-3800}

\author{Vitaliy~Fadeyev}
\affiliation{Santa Cruz Institute for Particle Physics, University of California at Santa Cruz, Santa Cruz, CA 95064, USA}

\author{R.~Clive~Field}
\affiliation{SLAC National Accelerator Laboratory, Menlo Park, CA 94025, USA}

\author{Alessandra~Filippi}
\affiliation{Istituto Nazionale di Fisica Nucleare, Sezione di Torino,  Torino, Italy}

\author{Sarah~Gaiser}
\affiliation{Stanford University, Stanford, CA 94305}

\author{Nerses~Gevorgyan}
\affiliation{Yerevan Physics Institute, 375036 Yerevan, Armenia}

\author{Norman~Graf}
\affiliation{SLAC National Accelerator Laboratory, Menlo Park, CA 94025, USA}

\author{Mathew~Graham}
\affiliation{SLAC National Accelerator Laboratory, Menlo Park, CA 94025, USA}

\author{Michel~Guidal}
\affiliation{Institut de Physique Nucleaire d'Orsay, IN2P3, BP 1, 91406 Orsay, France}

\author{Ryan~Herbst}
\affiliation{SLAC National Accelerator Laboratory, Menlo Park, CA 94025, USA}

\author{Maurik~Holtrop}
\affiliation{University of New Hampshire, Department of Physics, Durham, NH 03824}

\author{John~Jaros}
\affiliation{SLAC National Accelerator Laboratory, Menlo Park, CA 94025, USA}

\author{Robert~Johnson}
\affiliation{Santa Cruz Institute for Particle Physics, University of California at Santa Cruz, Santa Cruz, CA 95064, USA}

\author{Valery~Kubarovsky}
\affiliation{Thomas Jefferson National Accelerator Facility, Newport News, Virginia 23606}

\author{Dominique~Marchand}
\affiliation{Institut de Physique Nucleaire d'Orsay, IN2P3, BP 1, 91406 Orsay, France}

\author{Luca~Marsicano}
\affiliation{Istituto Nazionale di Fisica Nucleare, Sezione di Genova e Dipartimento di Fisica dell'Universit\`a, 16146 Genova, Italy}

\author{Takashi~Maruyama}
\affiliation{SLAC National Accelerator Laboratory, Menlo Park, CA 94025, USA}

\author{Samantha ~McCarty}
\affiliation{University of New Hampshire, Department of Physics, Durham, NH 03824}

\author{Jeremy~McCormick}
\affiliation{SLAC National Accelerator Laboratory, Menlo Park, CA 94025, USA}

\author{Bryan~McKinnon}
\affiliation{University of Glasgow, Glasgow G12 8QQ, United Kingdom}

\author{Omar~Moreno}
\affiliation{SLAC National Accelerator Laboratory, Menlo Park, CA 94025, USA}

\author{Carlos~Munoz-Camacho}
\affiliation{Institut de Physique Nucleaire d'Orsay, IN2P3, BP 1, 91406 Orsay, France}

\author{Timothy~Nelson}
\affiliation{SLAC National Accelerator Laboratory, Menlo Park, CA 94025, USA}

\author{Silvia~Niccolai}
\affiliation{Institut de Physique Nucleaire d'Orsay, IN2P3, BP 1, 91406 Orsay, France}

\author{Rory~O'Dwyer}
\affiliation{Stanford University, Stanford, CA 94305}

\author{Rafayel~Paremuzyan}
\affiliation{Thomas Jefferson National Accelerator Facility, Newport News, Virginia 23606}

\author{Emrys~Peets}
\affiliation{Stanford University, Stanford, CA 94305}

\author{Nunzio~Randazzo}
\affiliation{Istituto Nazionale di Fisica Nucleare, Sezione di Catania e Dipartimento di Fisica dell'Universit\`a, Catania, Italy}

\author{Benjamin~Raydo}
\affiliation{Thomas Jefferson National Accelerator Facility, Newport News, Virginia 23606}

\author{Benjamin~Reese}
\affiliation{SLAC National Accelerator Laboratory, Menlo Park, CA 94025, USA}

\author{Philip~Schuster}
\affiliation{SLAC National Accelerator Laboratory, Menlo Park, CA 94025, USA}

\author{Gabriele~Simi}
\affiliation{Istituto Nazionale di Fisica Nucleare, Sezione di Padova,  Padova, Italy}

\author{Valeria~Sipala}
\affiliation{Universit\`a di Sassari e Istituto Nazionale di Fisica Nucleare, 07100 Sassari, Italy}

\author{Matthew~Solt}
\affiliation{SLAC National Accelerator Laboratory, Menlo Park, CA 94025, USA}

\author{Alic~Spellman}
\affiliation{Santa Cruz Institute for Particle Physics, University of California at Santa Cruz, Santa Cruz, CA 95064, USA}

\author{Stepan~Stepanyan}
\affiliation{Thomas Jefferson National Accelerator Facility, Newport News, Virginia 23606}

\author{Holly~Szumila-Vance}
\affiliation{Thomas Jefferson National Accelerator Facility, Newport News, Virginia 23606}

\author{Lauren~Tompkins}
\affiliation{Stanford University, Stanford, CA 94305}

\author{Natalia~Toro}
\affiliation{SLAC National Accelerator Laboratory, Menlo Park, CA 94025, USA}

\author{Maurizio~Ungaro}
\affiliation{Thomas Jefferson National Accelerator Facility, Newport News, Virginia 23606}

\author{Hakop~Voskanyan}
\affiliation{Yerevan Physics Institute, 375036 Yerevan, Armenia}

\begin{abstract}

\newpage

The Heavy Photon Search (HPS) experiment is designed to search for a new vector boson \aprime in the mass range of \SI{20}{MeV/c^2} to \SI{220}{MeV/c^2} that kinetically mixes with the Standard Model photon with couplings $\epsilon^2 >10^{-10}$. In addition to the general importance of exploring light, weakly coupled physics that is difficult to probe with high-energy colliders, a prime motivation for this search is the possibility that sub-\si{GeV} thermal relics constitute dark matter, a scenario that requires a new comparably light mediator, where models with a hidden $U(1)$ gauge symmetry, a ``dark'', ``hidden sector'', or ``heavy'' photon, are particularly attractive. HPS searches for visible signatures of these heavy photons, taking advantage of their small coupling to electric charge to produce them via a process analogous to bremsstrahlung in a fixed target and detect their subsequent decay to \epem pairs in a compact spectrometer. In addition to searching for \epem resonances atop large QED backgrounds, HPS has the ability to precisely measure decay lengths, resulting in unique sensitivity to dark photons, as well as other long-lived new physics. After completion of the experiment and operation of engineering runs in 2015 and 2016 at the JLab CEBAF, physics runs in 2019 and 2021 have provided datasets that are now being analyzed to search for dark photons and other new phenomena.

\end{abstract}

%% file: introduction.tex
\label{sec:introduction}

\subsection{Physics Motivations and Signatures}
\label{sec:physics}

The last decade has seen growing interest in the search for new forces mediated by a \si{GeV}-scale or lighter force carrier, with weak coupling to ordinary matter (see e.g.~\cite{Hewett:2012ns,Essig:2013lka,Alexander:2016aln,Battaglieri:2017aum} and references therein).  There are two distinct motivations for such searches.  First, such a force could play an essential role in the physics of dark matter. Specifically, if the dark matter is ``light'', below about \SI{2}{GeV}, and if its thermal production during the early universe is to account for the present relic abundance, then it must be associated with a new light mediator and be part of a hidden sector. Evidence for a dark force could be the first compelling evidence for a hidden sector and lead to identifying the dark matter. Second, the systematic exploration of weakly coupled physics at low masses is an important complement to the search for new physics at the Large Hadron Collider (LHC) and other high-energy machines.  Owing to their low masses and weak couplings, new force carriers can be hard to identify in high-energy collider searches.  The discovery of such a new force would represent a profound shift in our understanding of the laws of Nature.

Heavy photons (also called ``dark'' or ``hidden-sector'' photons)  are a canonical example of a new force that could couple directly to the dark matter and also appear in many new-physics scenarios beyond the Standard Model (SM).  Kinetic mixing of the heavy photon with the SM photon is generically weak if the mixing arises from radiative corrections~\cite{Holdom:1985ag,Galison:1983pa}.  This gives rise to a commensurately weak coupling of the heavy photon to electrically charged particles, $\epsilon e$, where $e$ is the electron charge and $\epsilon \lesssim 10^{-2}$. Consequently, electrons can radiate heavy photons in a process analogous to bremsstrahlung at a rate suppressed by $\sim \epsilon^2 m_{\mathrm{e}}^2 / m_{\maprime}^2$, and heavy photons can decay to pairs of charged particles, as allowed by conservation laws~\cite{AprimeFixedTargetTheory}.  Over the past decade, searches for heavy photons have been conducted over large regions of the heavy photon mass/coupling parameter space, but much of that parameter space, including territory favored by thermal dark matter production in the early universe, remains unexplored. While the heavy photon model has been a primary motivation, these searches are also sensitive to more general models of dark forces with vector, axial-vector, scalar, or pseudo-scalar couplings to matter, since mediators of other spins have production and decay properties similar to that of heavy photons.

Since the production rate for heavy photons is strongly suppressed relative to analogous processes involving regular photons, experiments must always contend with an overwhelming, kinematically identical QED background. Two handles are available to distinguish a signal from this background. First and most obviously, decays of on-shell mediators result in a mass peak, and a number of existing and future experiments have sensitivity to sub-\si{GeV} dark photons at larger couplings ($\epsilon^2 \gtrsim 10^{-8}$) through searches for narrow \epem or $\mu^+\mu^-$ resonances atop a continuum background. Requiring high intensities to reach small couplings, these experiments require intense beams and typically spread out final state particles into large spectrometers to achieve reasonable occupancies, and reaching lower couplings through brute force is difficult, requiring orders of magnitude more luminosity. However, at smaller couplings and lower masses, dark photons become long lived and can travel macroscopic distances before decaying, with $\gamma c \tau \propto 1/{m_{\maprime}}^2\epsilon^2$, offering another possibility to eliminate backgrounds. In the extreme case --- at very small couplings and masses --- extremely high intensity, high energy beam dumps with a detector behind a shield and decay region offer sensitivity to very small couplings with relatively simple low-rate detectors. However, at intermediate couplings --- too small for simple spectrometers and too large for beam dumps where decay lengths are in the range of $\sim$\SIrange{1}{100}{mm} --- the key to sensitivity is the elimination of prompt backgrounds through high-purity reconstruction and identification of long-lived \aprime decays. In addition to having unique sensitivity to heavy photons, experiments with this capability have generic sensitivity to new light, weakly-coupled physics with long lifetimes, including axion-like particles (ALPs), strongly interacting massive particles (SIMPs), and inelastic dark matter (iDM).~\cite{Berlin:2018bsc}~\cite{Berlin:2018tvf}~\cite{iDM:2017}

\subsection{HPS Design Principles}
\label{sec:principles}

\begin{figure}[!htb]
 \centering
 \includegraphics[width=0.9\textwidth]{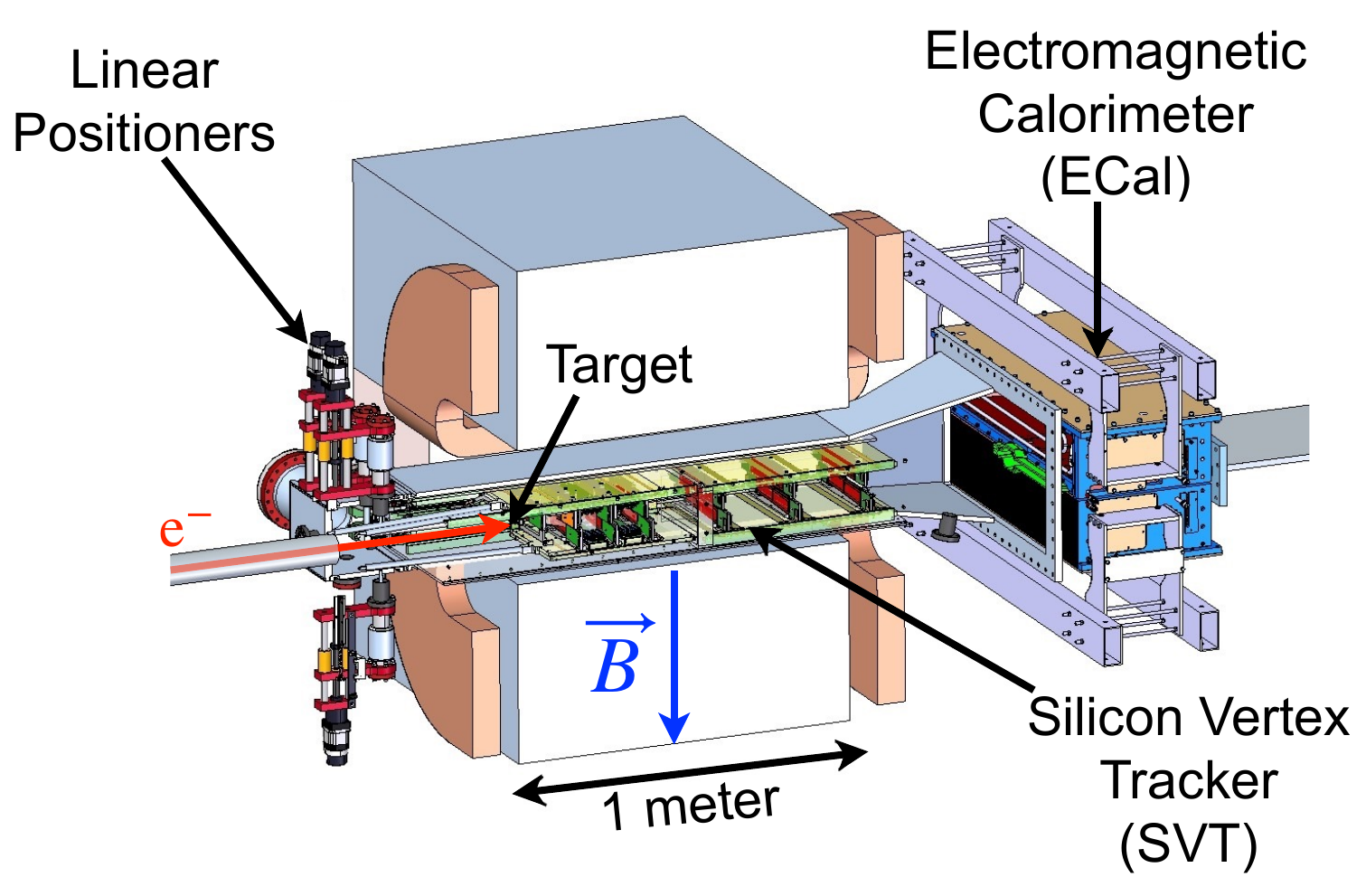}
 \caption{A cutaway view of the baseline HPS detector showing the Silicon Vertex Tracker (SVT) in a vacuum chamber inside the bore of the spectrometer magnet and the Electromagnetic Calorimeter (ECal) downstream.  The positions of the target and the front portions of the SVT are controlled by a set of linear positioning motors upstream of the detector.}
 \label{fig:detector}
\end{figure}

The Heavy Photon Search (HPS) experiment exploits these signatures --- reconstruction of both mass and decay length --- to search for dark photons and other new physics at the Thomas Jefferson National Accelerator Facility (JLab) in Newport News, Virginia. Shown in Figure~\ref{fig:detector}, HPS is a compact \epem spectrometer 
built inside of a standard dipole analyzing magnet.
The compact size of the experiment allows for large acceptance in a small and easily-sited footprint, and uses fast and highly granular detectors to provide triggering, tracking and vertexing, and particle identification by reusing inexpensive and readily available technologies developed for other projects.
HPS provides sensitivity to a range of \aprime masses by operating at a range of beam energies from $\sim$\SIrange{1}{6}{GeV} and uses intense electron beams (\SIrange{50}{500}{nA}) on thin (\SIrange{4}{20}{\um} foils) tungsten targets to maximize signal rates relative to QED backgrounds.

While rejection of QED backgrounds motivates the best possible resolutions for \epem mass and vertex position, the kinematic characteristics of the signal and beam backgrounds determine the overall layout of the HPS apparatus. Radiation of a mediator that is heavy compared to the incoming electron carries away most of the energy in the reaction, so $x = E_{\maprime}/E_\mathrm{beam}$ is peaked strongly at \num{1}. The relatively light mediator, being highly boosted, has its momentum closely aligned with the beam direction and releases relatively little energy in decaying to an \epem pair, leaving that pair also boosted in the very forward direction and azimuthally back-to-back with respect to the beam direction. Therefore, it is critical for the experiment to instrument the far forward region to search for \epem pairs on either side of the beam direction. Given the vertical magnetic field of the spectrometer dipole, a high rate beam electrons degraded by passage through the target are spread along the horizontal plane, so all of the detector subsystems are split above and below the beam plane. HPS employs a high-rate silicon tracking and vertexing detector (SVT) with sensors positioned as close as possible to the target and the plane of the through-going beam to provide the best possible acceptance around this horizontal ``sheet of flame''. The extent of the forward acceptance is limited by the background rate of single beam electrons that scatter in the target, which cannot mimic signal but create extreme occupancies ($\approx$\SI{10}{MHz/mm^2}) at the edge of the first silicon layer. As a consequence, the high repetition rate of the CEBAF beam (\SI{499}{MHz}), in tandem with a high-rate \epem trigger with precision timing and similarly precise timing in the SVT is required to select only in-time hits for reconstruction.  HPS uses a lead tungstate electromagnetic calorimeter (ECal) and a scintillator hodoscope to provide this trigger via high performance trigger and data acquisition hardware.

In addition to the kinematically irreducible background from radiative tridents and the nuisance occupancy of single scattered electrons that can spoil event reconstruction, there are two other important backgrounds to consider. The first are trident events from the Bethe-Heitler process, which differ from radiative events in being peaked at low $x$. While this distinction provides a critical handle to mitigate and estimate this contribution, the cross section for Bethe-Heitler tridents is so large that these events are still the dominant source of background near $x=1$. Second are events where a hard, wide angle, bremsstrahlung photon converts in the target or detector material and the resulting positron is paired with the recoil electron. This background also has a different kinematic signature that can be used to estimate its contribution.

%% file: experimental_overview.tex
\label{sec:experimental_overview}




The HPS experiment was proposed to the JLab PAC in 2011 and won approval for 180 days of operations, conditional upon demonstrating the ability to meet key technical challenges with a test run. A small scale demonstrator including all of the major subsystems --- SVT, ECal, trigger, and DAQ --- was built and operated in 2012, after which the project won full approval in 2013 and construction commenced.~\cite{testRunNim} The apparatus was completed and installed in 2015 in advance of a short commissioning and engineering run, followed by another similar run in 2016. While this baseline detector performed as designed, analysis of these first datasets motivated some key improvements and additions to the apparatus. In particular, the initial design studies had overestimated the acceptance of the trigger and SVT. In response, an additional layer was added to the SVT, even closer to the target, and a positron hodoscope was added to the trigger system to cover a hole in the ECal necessitated by the high rate of scattered electrons in that region. With these improvements, HPS had first physics operations in 2019, followed by another run in fall 2021. The following sections outline the critical elements and evolution of the detector design.

\subsection{CEBAF, Beamline, and Target}

\begin{figure}[!htb]
 \centering
 \includegraphics[width=\textwidth]{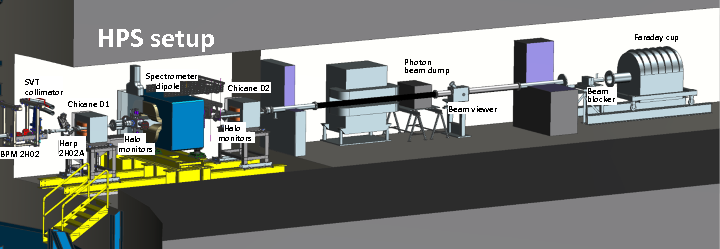}
 \caption{The HPS detector as installed as part of a three magnet chicane in the downstream alcove of Hall~B at the JLab CEBAF.}
 \label{fig:beamline}
\end{figure}

The HPS experiment operates in the downstream alcove of Hall~B of the Continuous Electron Beam Accelerator Facility (CEBAF) at JLab, as shown in Figure~\ref{fig:beamline}.
CEBAF is oval shaped, consisting of two linacs connected by a pair of recirculuating arcs, which enable injected beam to make multiple passes of the linacs --- gaining \SI{2.2}{GeV} per pass for up to \num{5.5} passes --- before extracting the beam into one of four halls. Sub-harmonics of the \SI{1.497}{GHz} beam may be simultaneously extracted into the different halls, allowing simultaneous operation of multiple experiments with high-rate (typically \SI{499}{MHz}) beam.~\cite{CEBAF}

Operation at the JLab CEBAF is fundamental to the success of the HPS experiment. First, the experiment requires a very high repetition rate multi-\si{GeV} electron beam with low per-bunch charge, together with precision hit timing in all subsystems, in order to screen out the high rate of background hits from scattered single electrons. A higher per-bunch charge would spoil the clean tracking and vertexing required for the displaced vertex search, while a lower current would require unacceptably long operations.  No other currently operating facility in the US can provide the required beam for the experiment. Meanwhile, the extraordinary proximity of the SVT layers to the beam, as close as \num{500} microns between the edges of sensors and the center of the beam, requires a very small beamspot ($<$\SI{50}{\um} vertically) with vanishing low halo rate ($\lesssim${}$10^{-6}$ outside the gaussian core) and excellent beam stability ($<$\SI{30}{\um} vertical variation). Ensuring the safety of the SVT also requires multiple diagnostic and protection systems. Diagnostics include beam position monitors, wire scanners, and beam halo monitors used to establish the trajectory and final focus of the beam and assure clean beam delivery during operations. Protection systems are both active --- a fast shutdown system fed by the halo monitors --- and passive --- a collimator with selectable aperture.~\cite{Baltzell:2016eee}

The target for the experiment is chosen to be as thin as possible given the upper limit on beam currents in Hall B to achieve the desired luminosity, in order to reduce two-step QED backgrounds in the target, like the converted WABs mentioned in Section~\ref{sec:introduction}. The target itself is a movable assembly with different thickness tungsten foils, ranging from \num{0.125}\% -- \num{0.625}\%~$X_0$, where the engineering runs also included carbon and polyethylene targets for calibration purposes.

\subsection{Silicon Vertex Tracker}

The SVT is the cornerstone of the HPS experiment, responsible for measuring both the mass and vertex position of \epem pairs. A number of competing requirements shape the design of the SVT. First, \aprime decay products have typical momenta $\lesssim E_\mathrm{beam}/2$, so multiple scattering dominates mass and decay length errors for any feasible material budget. Therefore, the construction of the SVT must place the smallest amount of material possible in the tracking volume. Second, the signal yields for long-lived \aprime are very small, so the rejection of prompt vertexes must be exceedingly strong, better than $10^{-6}$, to reduce prompt background to the order of one event or less. Finally, as previously discussed, the passage of scattered and degraded primary beam through the apparatus creates a region of extreme occupancy and radiation that is critical for sensitivity to low-mass \aprime that have decay products nearly collinear with the beam. This puts low-mass acceptance at odds with tracking and vertexing purity and the longevity of the detector, requiring careful design allow the largest usable acceptance.

The SVT employs conventional silicon microstrip sensors, which allows the readout and cooling material to be placed outside the tracking volume. The sensors are cooled from the ends to below \SI{-10}{\degreeCelsius} to extend their lifetime at peak fluences exceeding $10^{16}$~\eminus (or \num{4e14}[1~\si{MeV}~neq]). Because the regions of high occupancy are small spots in two dimensions, only a very short length of the edge strips see high occupancy, so even long strips covering those regions have occupancies only a small factor larger than what pixels would experience. To further reduce occupancies for tracking, the CMS APV25 chip is used for readout in ``multi-peak'' mode, which records multiple samples of the signal development, allowing reconstruction of hit time with $\approx$\SI{2}{ns} resolution --- near the level required to tag events in individual CEBAF bunches. Finally, to eliminate displaced events and occupancy from beam-gas collisions, the SVT must operate inside the beam vacuum.

\begin{figure}[!htb]
    \centering
    \includegraphics[width=0.49\textwidth]{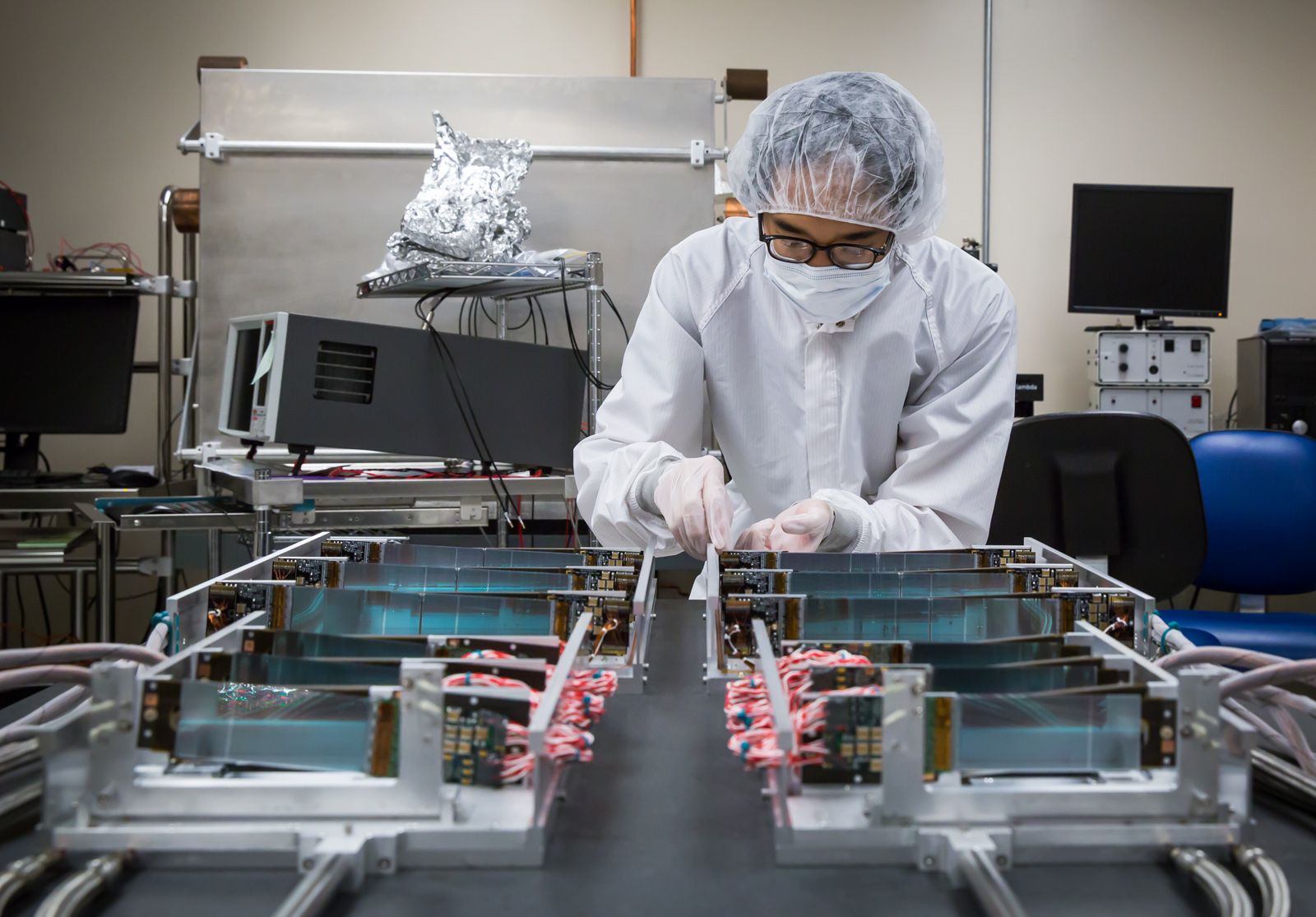}
    \hfill
    \includegraphics[width=0.50\textwidth]{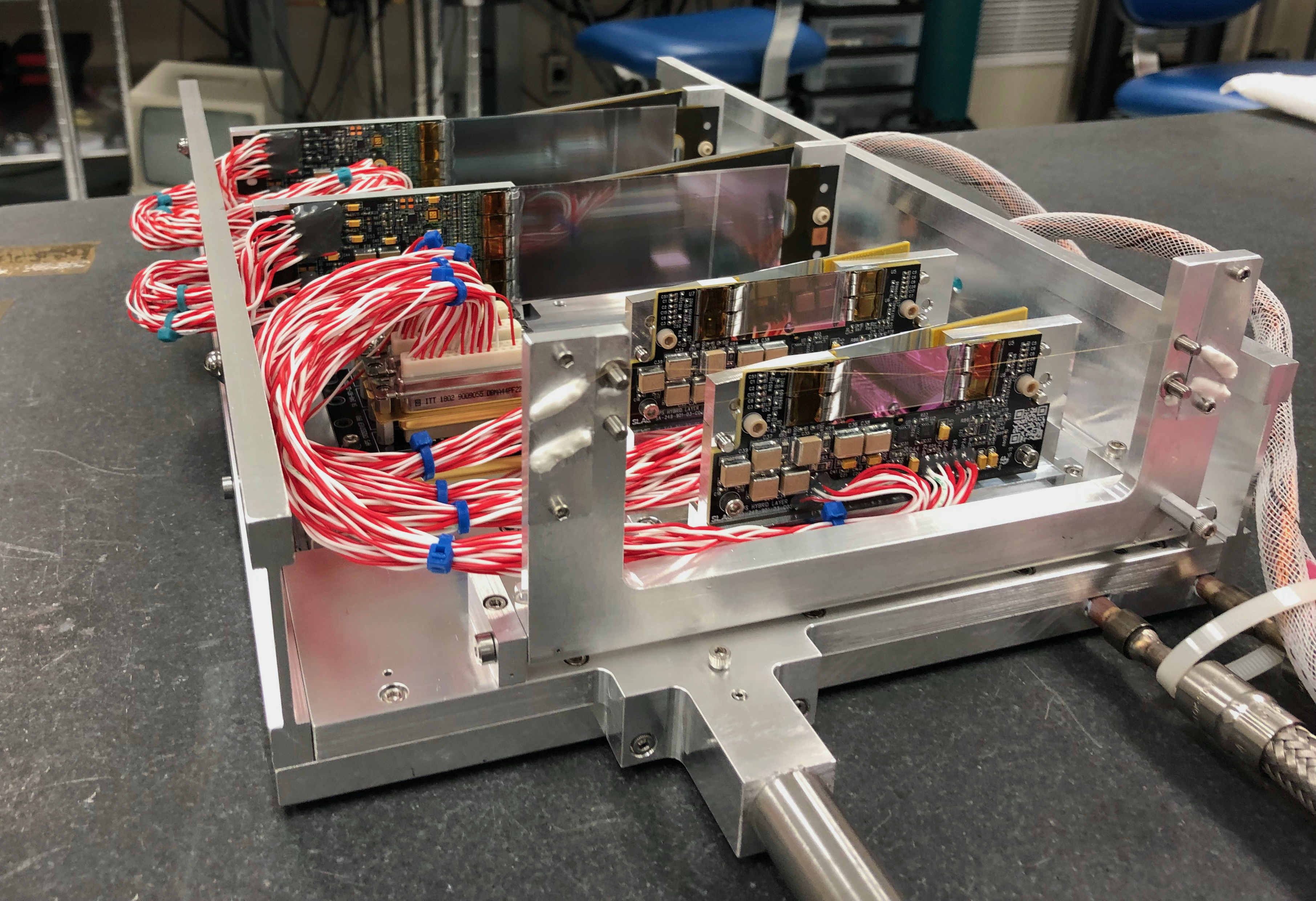}
    \caption{At left, the two halves (top and bottom) of the baseline SVT prior to installation at JLab in 2014. At right, front layers of the upgraded detector installed in 2019, including thinner, slim edge silicon sensors in the first two layers.}
    \label{fig:SVT}
\end{figure}

Shown in the left panel of Figure~\ref{fig:SVT}, the baseline SVT installed in 2015 consisted of 6 double-layer 3-d measurement stations above and below the beam plane arranged between $z=\SI{10}{cm}$ and $z=\SI{90}{cm}$ downstream of the target, where each measurement station comprised a stereo pair of individual sensors. 
The arrangement of these stations covered the forward region down to \SI{15}{mrad} on either side of the horizontal beam plane for prompt particles from the target. With the first layer at $z=\SI{10}{cm}$, the edge of the active region (silicon sensor) in the first layer was only \SI{1.5}{mm} (\SI{0.5}{mm}) from the beam plane. To allow for tuning of the beam through the detector, the SVT uses linear positioners to open and close around the beam plane.

While this baseline detector performed as designed, analysis of engineering run data motivated modifications to the design to improve resolution and acceptance. A seventh double-layer was added to the SVT at $z=\SI{5}{cm}$, using thin silicon to reduce material, slim edge processing to maintain the \SI{15}{mrad} acceptance, and short striplets split at the beamspot to halve the peak occupancy.~\cite{FADEYEV2020163991} This module design was also used to replace the previous first layer to improve resolution and allow it to be positioned closer to the beam, and other layers were similarly moved closer to the beam as occupancies allowed to increase forward acceptance for displaced decays. This upgraded configuration of the front layers of the SVT, shown in the right panel of Figure~\ref{fig:SVT}, was completed and installed in advance of the first physics run in 2019, and operated again in 2021.

\subsection{Electromagnetic Calorimeter}

The HPS Electromagnetic Calorimeter (ECal) plays two critical roles.~\cite{ECal:2017} First, it provides a trigger for \epem pairs with sufficient energy and time resolution to eliminate the overwhelming background of scattered single beam electrons. Second, it provides positive identification of electrons and positrons offline, with sufficient timing to tag their energy deposits to a single CEBAF bunch, which can then be used to demand coincidence in the SVT. Like the SVT, the ECal must contend with extremely high rates and be relatively radiation tolerant in order to match the angular acceptance of the SVT as closely as possible.

\begin{figure}[!htb]
    \centering
    \includegraphics[width=0.95\textwidth]{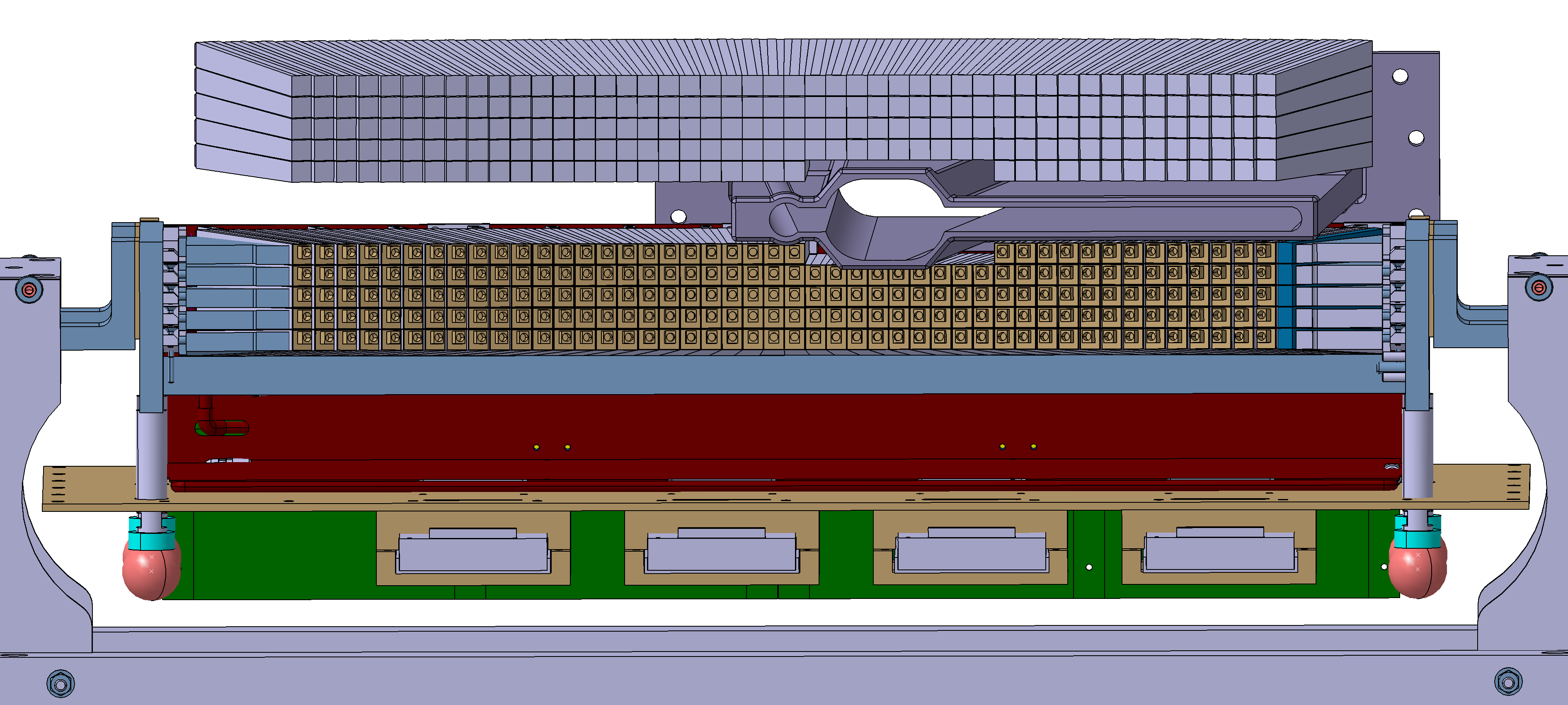}
    \caption{The crystal layout in the ECal, looking downstream along the beam direction. For clarity, only the crystals are shown in the top half, where the bottom includes  mechanical elements such as the motherboards (in green) and the copper plates for heat shielding (in red). Between the two halves of the ECal, the vacuum vessel extends toward the electron side to accommodate beam particles having lost energy in the target. The aperture is enlarged and 9 crystals are removed from each half, where occupancies would be unacceptably high from scattered beam.}
    \label{fig:HPS_ECal}
\end{figure}

The ECal meets these requirements through the use of 442 $\ce{PbWO4}$ crystals arranged in two identical arrays --- placed symmetrically above and below the beam plane downstream of the SVT. The through-going degraded beam is transported between the two halves in a vacuum chamber to eliminate beam-gas backgrounds. Each half is a matrix of 5x46 $\ce{PbWO4}$ crystals. From the first row of each half, 9 crystals are removed nearest the through-going beam as the rate of scattered beam electrons is intolerably high in that region, well in excess of \SI{1}{MHz}. The crystal layout and some mechanical elements of ECal are shown in Fig.\ref{fig:HPS_ECal}.

\subsection{Positron Hodoscope}

As noted in the previous section, nine modules from the rows of crystals nearest to the beam for each of the top and bottom ECal halves were removed due to unacceptably high rates from scattered beam electrons. After the engineering runs, it was found that the resulting hole in the acceptance allows up to half of electrons from $A^\prime \to \mathrm{e}^+ \mathrm{e}^-$ decay to escape detection. With the baseline ``pair trigger" -- requiring hits in the ECal for both particles -- events with an \eminus lost in the ECal hole, but otherwise tracked in SVT, were not recorded. To recover these events, a single-arm, positron trigger was implemented in advance of the first physics run in 2019. The rate of positrons is not high, but the positron side of the ECal is flooded with photons from bremsstrahlung in the target. For a single-arm positron trigger to work with an acceptable rate, the HPS detector was instrumented with a scintillation hodoscope covering the positron side of the ECal to distinguish electrons from photons. 

\begin{figure}[!htb]
 \centering
    \includegraphics[width=0.7 \textwidth]{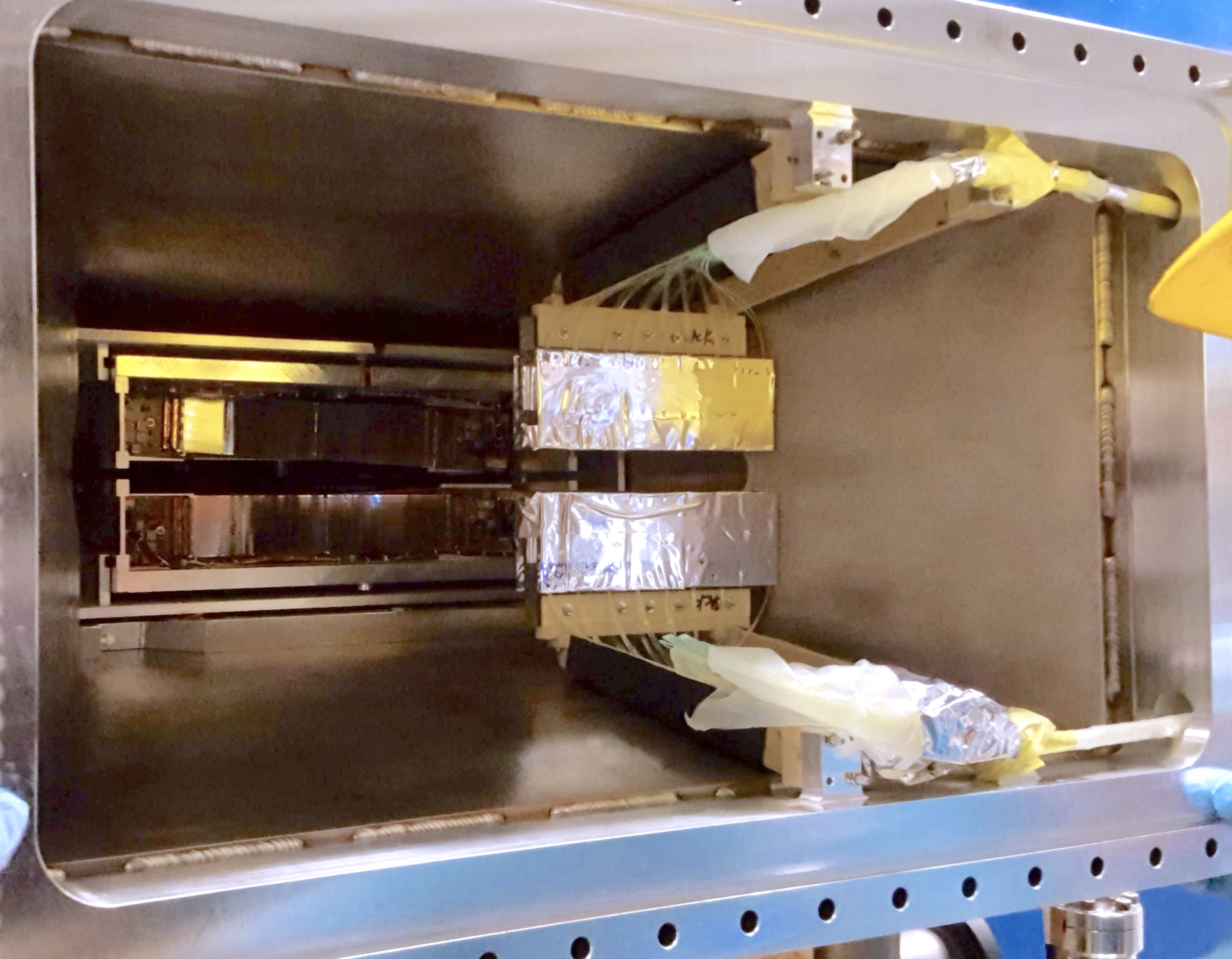}
    \caption{The positron hodoscope shown installed behind the longer modules in the back half of the SVT.}
    \label{fig:hodoscope}
\end{figure}

The hodoscope is installed inside the vacuum chamber between the SVT and the vacuum flange in front of the ECal as shown in Figure~\ref{fig:hodoscope}.
Each half of the hodoscope consists of two layers of overlapping scintillator tiles that match the vertical acceptance of the SVT, down to \SI{15}{mrad} above and below the beam plane.

\subsection{Trigger and DAQ}

As discussed in Section~\ref{sec:principles}, \aprime are produced nearly collinear with the beam so that the \epem decay daughters are typically back-to-back relative to the beam direction. Meanwhile, the vertical magnetic field of the spectrometer magnet will bend the electron and positron in opposite horizontal directions. Therefore, the primary trigger for the experiment is a ``pair trigger" in the ECal which requires energetic clusters in both halves (top and bottom) of the ECal, and with the clusters displaced horizontally from the centerline according to their energy, since lower-energy particles will curve more in the magnetic field. 

As previously discussed, the \eminus from some \aprime events are missed by the ECal due to the high-occupancy crystals removed from the design. The positron hodoscope installed before the 2019 Run recovers these events by requiring a cluster on the positron side only, again with an energy corresponding to horizontal position, in coincidence with a hodoscope in the same region.

The ECal and hodoscope are read out via VME-based \SI{250}{MHz} Flash ADC boards, capable of online tagging of triggered events to a \SI{4}{ns} window within a latency of less than \SI{3}{\us}. The trigger signal is passed to the SVT DAQ, based on the SLAC ATCA-based RCE platform, which initiates six-sample readout of the APV25 pipelines at \SI{41.667}{MHz}~\cite{RCE:7431254}. The SVT DAQ and ECal DAQ pass their data to PC-based event building to write the data to storage.  Typical trigger rates for HPS are \SIrange{20}{40}{kHz}, with data rates of \SIrange{250}{500}{MB/s}.

%% file: status_plans.tex
\label{sec:status_plans}

HPS is currently in the heart of its experimental campaign to search for dark photons and other dark sector phenomena. From the initial conceptual design roughly a decade ago, the baseline apparatus was first operated in 2015 and physics operations commenced in 2019 after improvements were made to increase the sensitivity of the experiment as described in Section~\ref{sec:experimental_overview}. The following sections outline the operational history of the experiment, the datasets collected to date, and the status of results from the experiment, as well as the expectation for future results and plans for future operations.

\subsection{Operational History and Datasets}

As discussed in Section~\ref{sec:experimental_overview}, two engineering runs were completed in 2015 and 2016 with the baseline HPS detector, and two physics runs were completed in 2019 and 2021 with an upgraded detector. The parameters of these runs and datasets are summarized in Table~\ref{tab:runs}.

\begin{table}[!hb] 
\setlength{\tabcolsep}{12pt}
    \centering
    \begin{tabular}{lcccc}
        \toprule
        \textbf{Run} & \textbf{Energy (\si{GeV})} & \textbf{Target (\%$\boldsymbol{X_0}$ W)} & \textbf{Beam Time Used} & $\boldsymbol{\int\mathcal{L}}$
        \textbf{pb}$\boldsymbol{^{-1}}$ \\
        \midrule
        2015 & 1.056 & 0.125 & 9.5 days & 1.17 \\
        2016 & 2.30 & 0.125 & 5.5 days & 10.75 \\
        2019 & 4.55 & 0.25/0.625 & 30 days & 122 \\
        2021 & 3.74 & 0.625 & 28 days & 168 \\
        \bottomrule
    \end{tabular}
    \caption{A summary of HPS runs and datasets to date. During periods of normal CEBAF operation, the overall combined efficiency of the accelerator and the HPS experiment is roughly 50\%. The 2015 and 2016 engineering runs included a larger fraction of commissioning time in addition to physics operations. After the 2021 Run, HPS is approved for 107 more days of operations.}
    \label{tab:runs}
\end{table}

In Spring 2015, HPS operated on nights and weekends over \SI{2.5}{weeks} at a beam energy of \SI{1.056}{GeV}. After a period of beam and detector commissioning and studies, a small physics dataset was collected with which to develop the analysis techniques for the experiment and perform a first search for heavy photons.
In Spring of 2016, data was collected on weekends over a \SI{10}{week} period at a beam energy of \SI{2.3}{GeV}. The detector was successfully commissioned much more quickly than in 2015, resulting in a significantly larger dataset with which to search for new phenomena.
In the summer of 2019, data was collected during dedicated operation in Hall~B over a period of \SI{11}{weeks} at a beam energy of \SI{4.55}{GeV}. Despite significant operational difficulties with CEBAF during this period, HPS successfully collected the first large physics dataset for the experiment.
Finally, after repairs to the SVT necessitated by damage sustained during 2019 operations, data was collected in the fall of 2021 during dedicated operation in Hall B over 8~weeks at a beam energy of \SI{3.74}{GeV}. The 2019 and 2021 datasets are large enough to provide significant sensitivity in the long-lived dark photon search, as will be discussed in Section~\ref{sec:reach}

\subsection{Summary of Previous Results}

The first HPS physics publication described the resonance search with the small dataset from the 2015 engineering run.~\cite{adrian2018search}.  This result is shown on the reach plots in Figure~\ref{fig:reach_future}. Because the 2015 dataset was far too small to allow for significant reach for long-lived \aprime, there was no corresponding publication for the displaced \aprime search, which has been reported in theses and conference proceedings~\cite{OmarThesis}~\cite{ShoThesis}~\cite{HollyThesis}~\cite{Moreno:2018tlx}. The resonance search on the 2016 dataset was completed in 2020, and has awaited completion of the displaced vertex analysis in 2021 for a combined publication of these results as a longer review article, currently in draft. Multiple theses have reported preliminary results on the 2016 data as analysis techniques have developed.~\cite{SebouhThesis}~\cite{SoltThesis}

\subsection{Projected Sensitivities and Future Prospects}
\label{sec:reach}

\begin{figure}[!htb]
    \centering
    \hfill
    \includegraphics[width=0.48\textwidth]{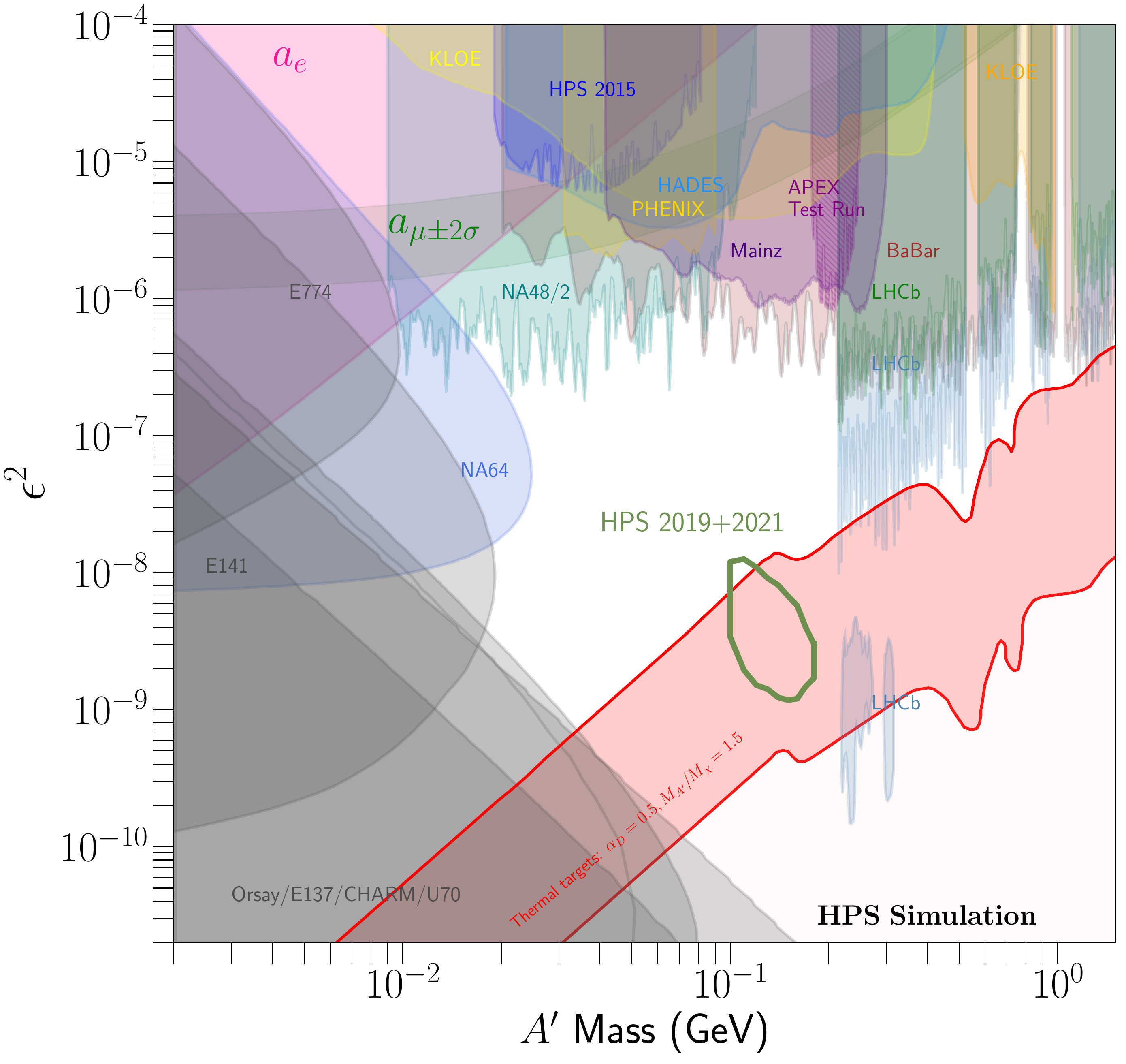}
    \hfill
    \includegraphics[width=0.49\textwidth]{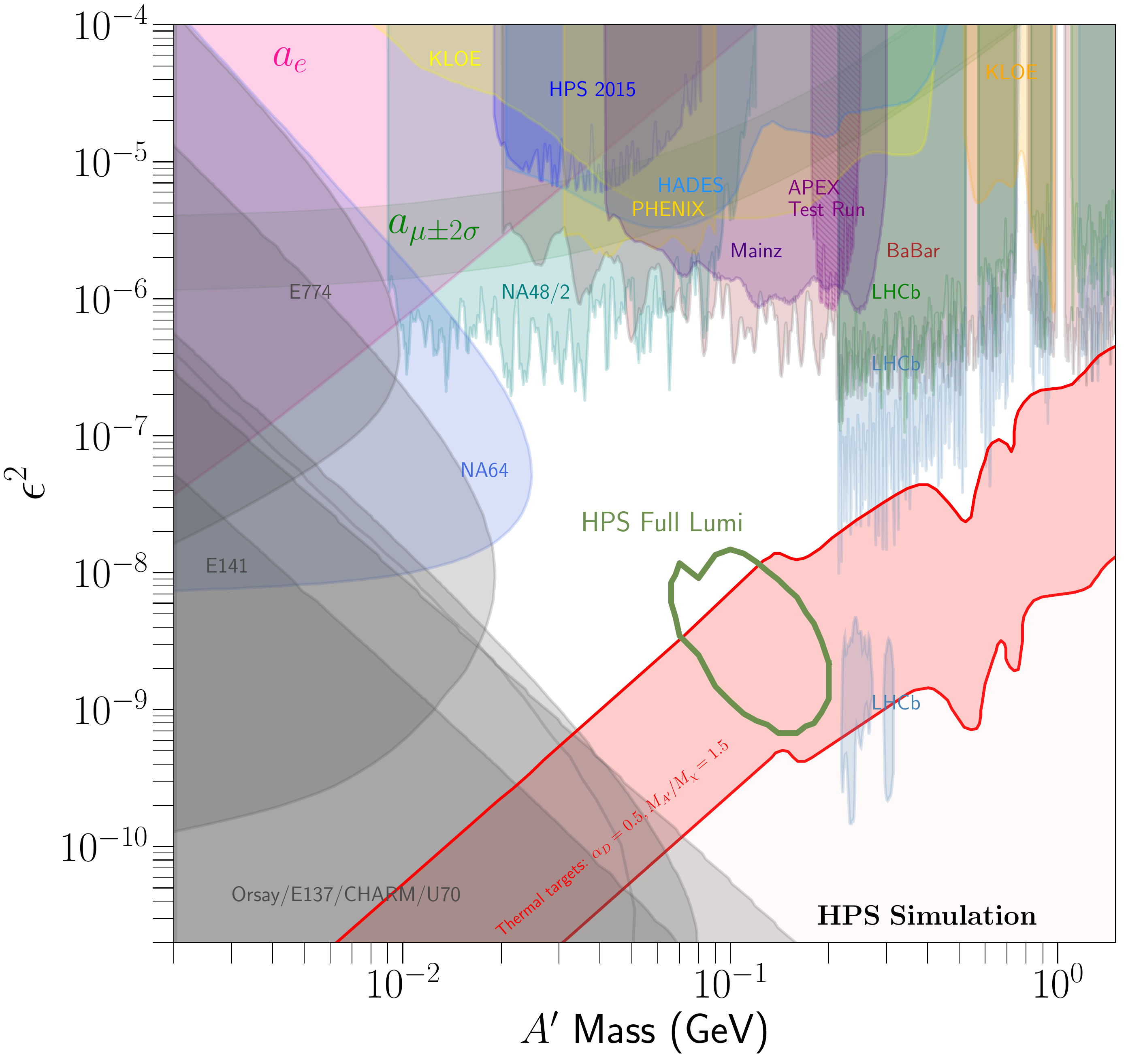}
    \hfill
    \caption{The reach anticipated from the displaced vertex analysis on the 2019 and 2021 datasets (left, green contour) and 
             proposed future operations (right, green contour) that utilize the remainder of the approved running time
a             for HPS. Existing limits from beam dump, collider and fixed target experiments are also shown along with regions favored and excluded by measurements of the anomalous magnetic moments of the muon and electron respectively. The pink band denotes the range of so-called thermal targets for freeze-out dark matter through a dark photon where a comprehensive review, including all exclusions, can be found in \cite{Battaglieri:2017aum, LHCb:2019vmc,NA64:2019auh}.}
    \label{fig:reach_future}
\end{figure}

Careful and complete analysis of the engineering run datasets have fostered a number of key improvements to the analysis techniques and apparatus, and provide a solid foundation for reach estimates with future datasets. Using the actual performance of these techniques on engineering run data as a benchmark, the expected reach of the HPS displaced vertex search with the combined 2019 and 2021 datasets is shown in the left panel of Figure~\ref{fig:reach_future}.

In considering the future, the HPS experiment was approved by the JLab PAC for \SI{180}{days} of operations. Subsequent to operations in 2021, the experiment has 107 PAC-approved days remaining for datataking. Under the assumption of two more periods of operations, split roughly equally between energies of \SI{2}{GeV} and \SI{4}{GeV}, and with a week of commissioning before each, the reach of the displaced vertex search, unique in its ability to probe highly motivated parameter space for thermal relics, is shown in the right panel of Figure~\ref{fig:reach_future}.


In addition to the minimal \aprime model, HPS has sensitivity to other physics scenarios.  These include axion-like particles (pseudo-scalars) with electron couplings, models with strong dynamics in the dark sector (SIMPs) and inelastic dark matter (iDM) scenarios. In some cases (e.g. ALPs), a straightforward re-interpretation of the dark photon search may suffice, but other cases (e.g. SIMPs and iDM, which involve missing energy as part of the signature) motivate targeted changes to the analysis strategy to maximize reach. It is expected that HPS has unique sensitivity to all of these possibilities, as well as generic models that include new, light, weakly coupled physics, especially in parts of parameter space with long-lived two-body decays. A search for SIMPs is already underway by the experiment using the well-understood data from the 2016 engineering run, and an analogous iDM search is under investigation, where these searches may already have groundbreaking sensitivity with the 2016 data. In addition to searching for new physics motivated by specific models such as dark photons, ALPs, SIMPs, and iDM, HPS is exploring the possibility of casting results in terms of a generic search for long-lived particles, to allow re-interpretation of results as new theoretical motivations are developed.



\subsection{Possible Extensions of the HPS Concept}

While HPS is well optimized to achieve unique sensitivity to dark photons with a compact and relatively inexpensive apparatus, the technologies used by the experiment could be employed in a different configuration or on a larger scale to cover other parts of the mass-coupling parameter space. Studies initiated during the previous Snowmass exercise explored two such possibilities.

\begin{figure}[!htb]
    \centering
    \includegraphics[width=0.44\textwidth]{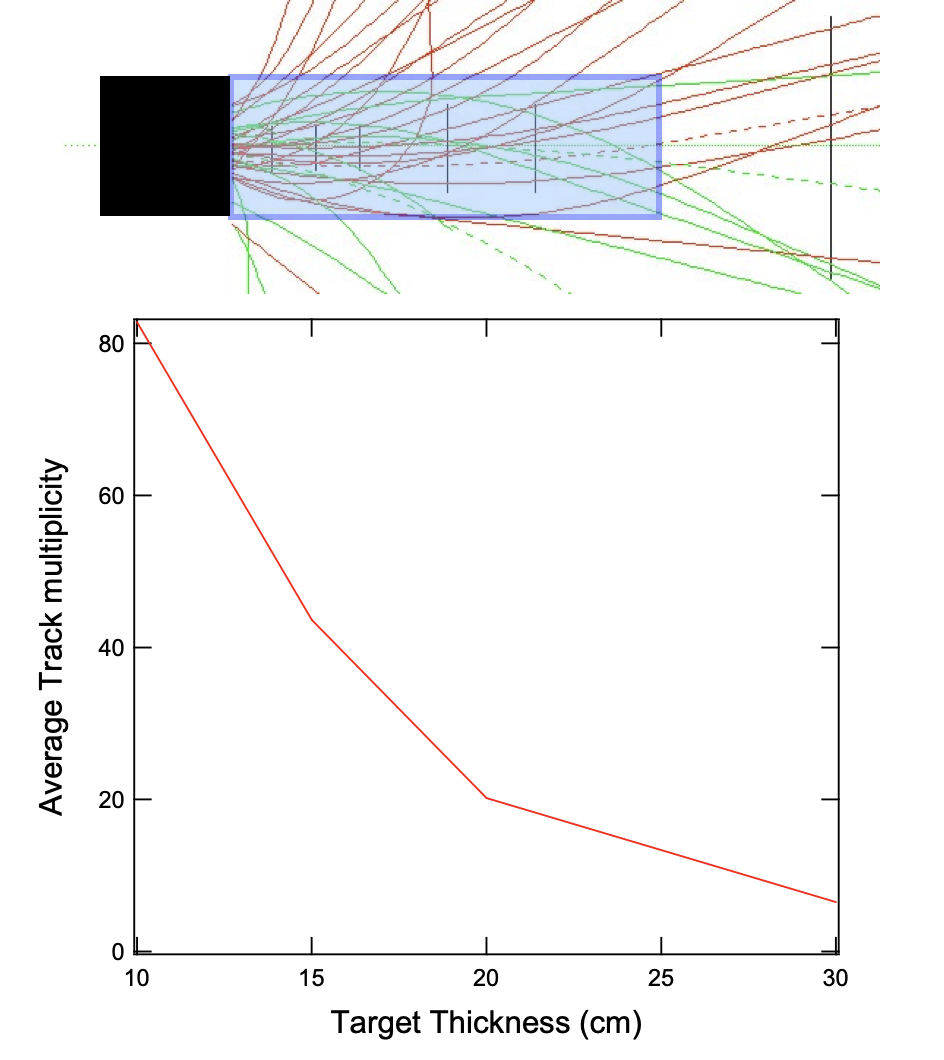}
    \hfill
    \raisebox{0.00\height}{
    \includegraphics[width=0.54\textwidth]{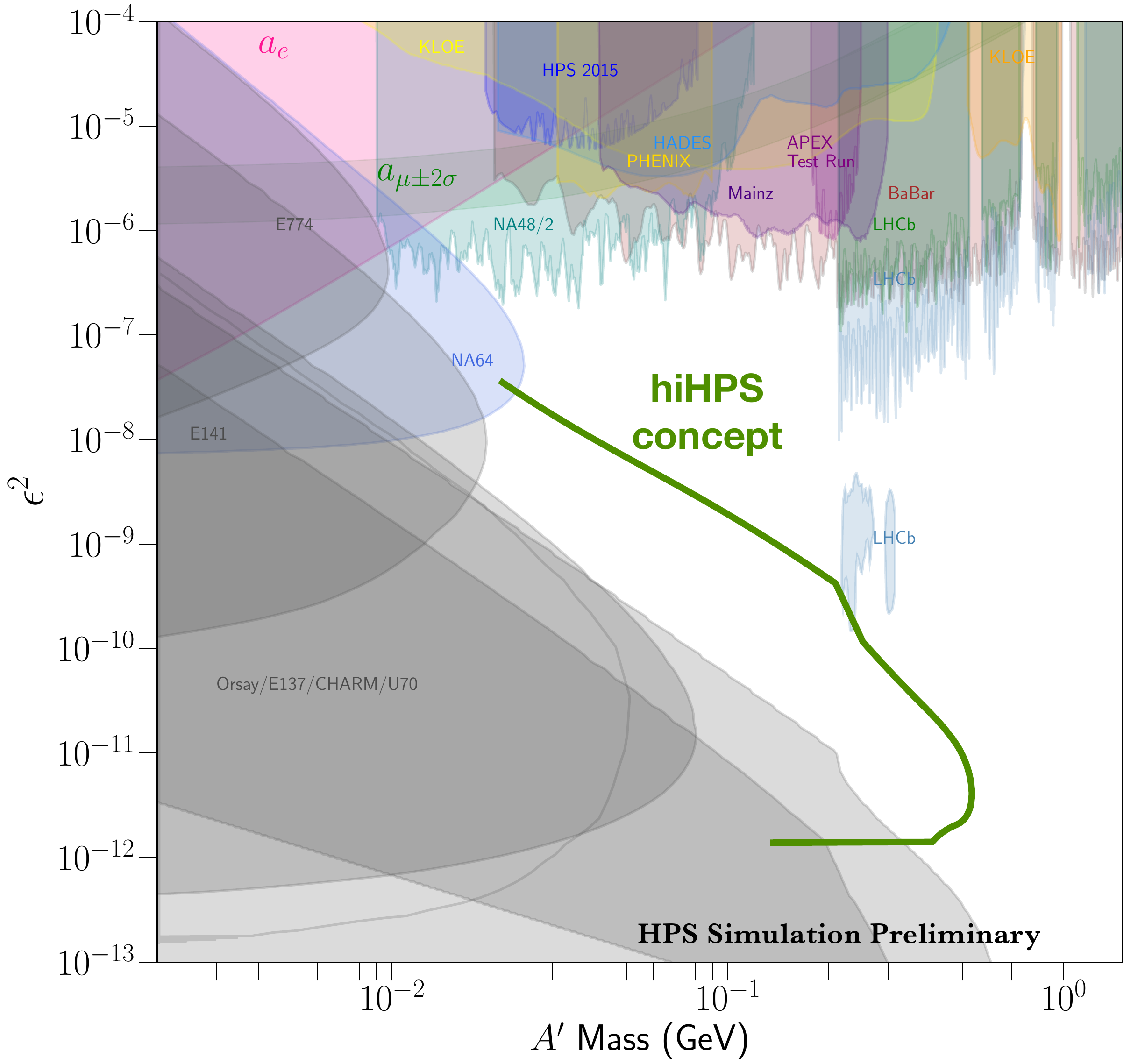}}
    \caption{At left, the charged-particle occupancy within an \SI{8}{ns} window (relevant to tracking with HPS SVT hardware) as a function of the thickness of a tungsten target and shield for the hiHPS concept, along with a representative simulated background event showing the distribution of positively (negatively) charged particles in green (red). The black region represents the tungsten target/shield and the light blue region is the bore of the dipole magnet with the HPS SVT inside. At right, a reach estimate (green line) for operation of the hiHPS concept as discussed in the text, assuming \num{0.5} events of expected background within each mass window.}
    \label{fig:hiHPS}
\end{figure}
The first of these concepts imagines placing the HPS apparatus behind the thinnest shield capable of screening out backgrounds for the \epem final state, which appears achievable with a \SI{30}{cm} thick tungsten target and shield with a \SI{6.6}{GeV} beam.  In this shallow dump configuration, shown in Figure~\ref{fig:hiHPS}, the detector would have no dead zone between the top and bottom halves, and the experiment would be run with a current of \SIrange{1}{10}{\uA} to allow \SIrange{20}{25}{C} of charge on target within a \SIrange{1}{10}{month} period at the limits of radiation dose in the tracker. 
The signature for the search is an \epem pair at a given mass, originating from a common vertex downstream of the tungsten shield, where the momentum of the pair points back from the vertex to the beamline inside the volume of the tungsten. The key challenges for this concept, termed high intensity HPS (hiHPS), are the cooling for the target (\SIrange{6.6}{66}{kW}) and the extreme flux of fast forward-going neutrons generating large fluences in the tracker and calorimeter. The target can likely be made feasible somewhere in this intensity range by surrounding the tungsten core with a cooled copper jacket, but the neutron dose appears more difficult, especially for the electromagnetic calorimeter required to distinguish electrons from pions, protons, and muons, likely requiring a much more aggressive technology than the lead-tungstate used by HPS. While much more extensive simulation of this concept would be required to verify that a near zero background search is feasible, the left panel of Figure~\ref{fig:hiHPS} gives an indication of the potential such a spectrometer behind a shallow electron beam dump in searching for dark photons at small masses and couplings. Not surprisingly, it is similar at low masses to the sensitivity provided by spectrometers behind shallow high-energy proton beam dumps.~\cite{2021_DarkQuest}

\begin{figure}[!htb]
    \centering
    \includegraphics[width=0.48\textwidth]{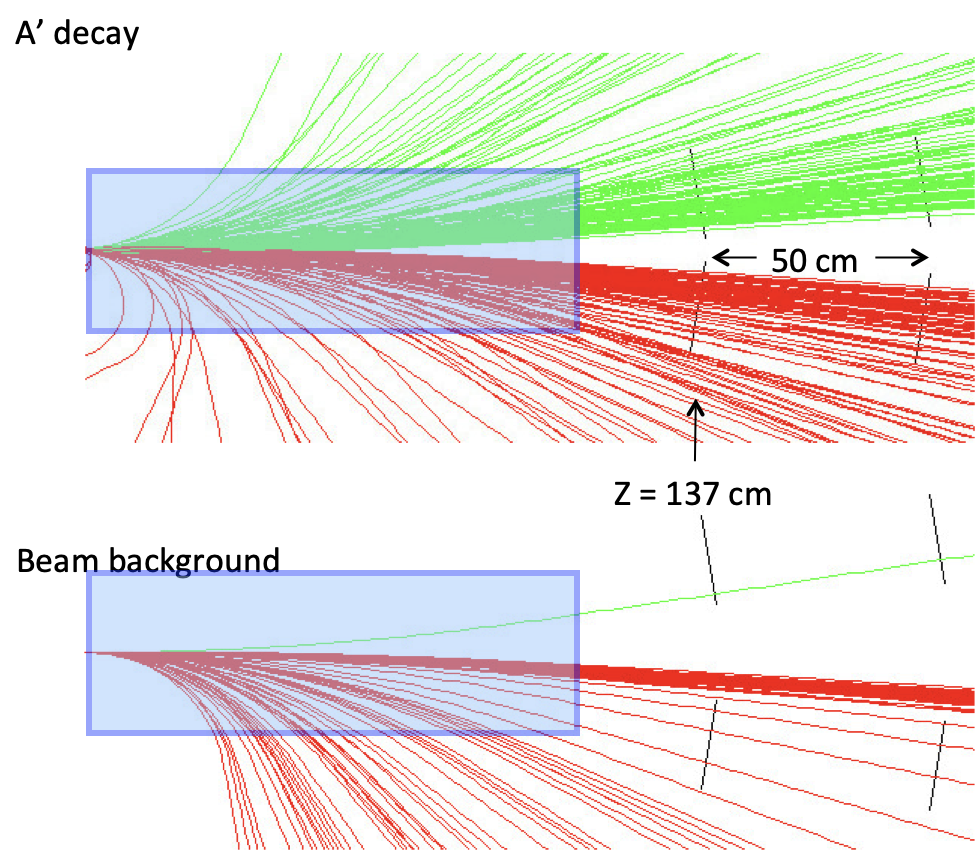}
    \hfill
    \raisebox{-0.05\height}
    {\includegraphics[width=0.46\textwidth]{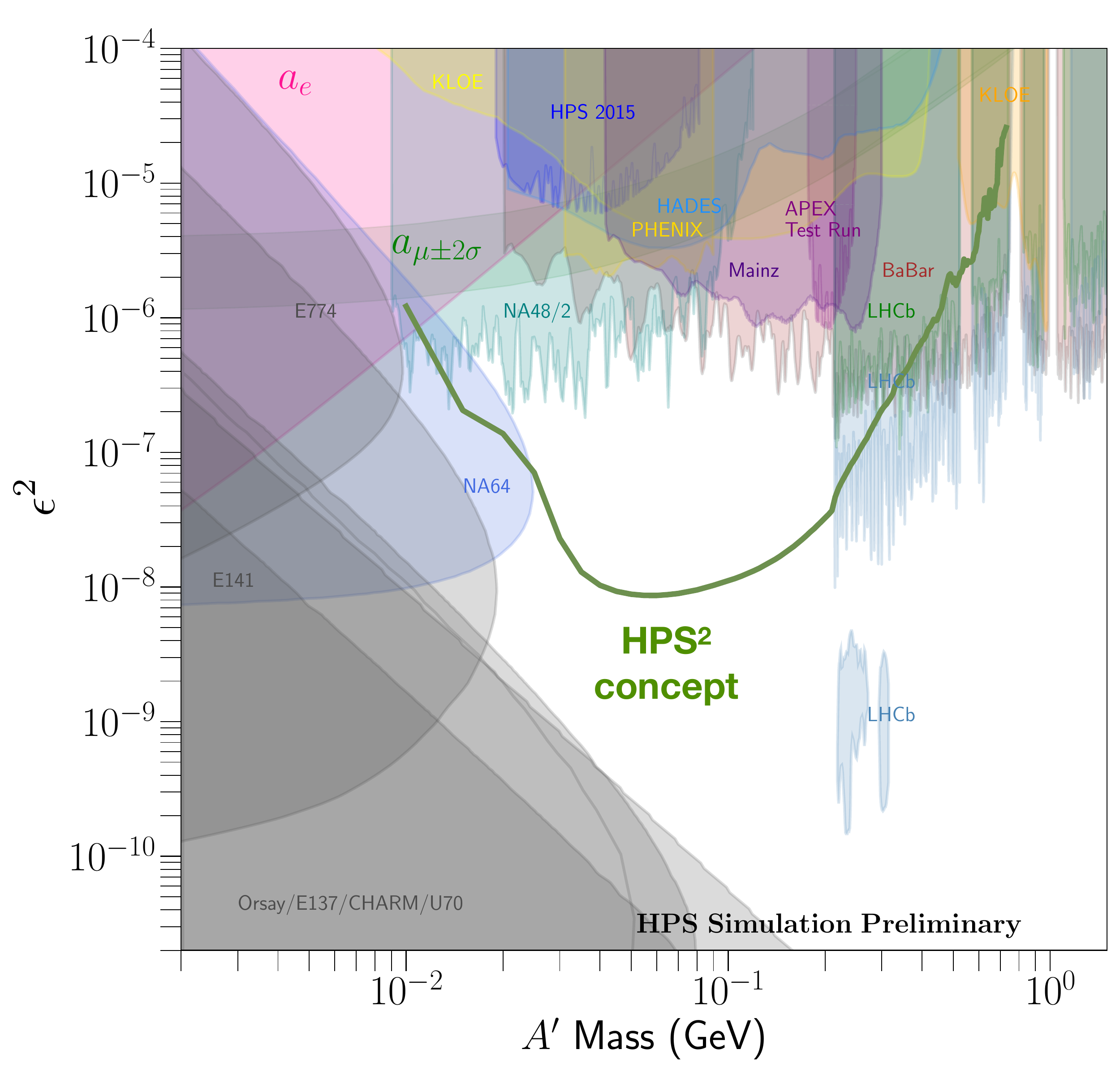}}
    \caption{At left, the distribution of \epem pairs in dark photon signal events for $M_\maprime=\SI{50}{MeV}$ and the distribution of charged particles from beam background for the HPS$^2$ concept. Only positively (negatively) charged particles are shown in green (red). The light blue region is the bore of the dipole magnet and the first two planes of a two-armed silicon tracker using the longer modules from the HPS SVT are shown downstream. At right, a reach estimate (green line) for operation of the HPS$^2$ concept as discussed in the text.}
    \label{fig:SuperHPS}
\end{figure}

The second of these concepts imagines placing a pair of HPS-like trackers behind the same spectrometer dipole used by HPS in a configuration reminiscent of a compact two-armed spectrometer, for a high acceptance, high rate resonance search.~\cite{Abrahamyan:2011gv} With the target still at the front face of the magnet as in HPS, the electron-positron pairs are well separated in the two arms of the tracker, as shown in Figure~\ref{fig:SuperHPS}. 

In this concept, termed HPS$^2$, there would be no dead zone in the positron side, which would have an eletromagnetic calorimeter for triggering and particle identification. Meanwhile, the photons and through-going beam are well separated from the electron arm, so the dead zone on the electron side can be much smaller than in HPS, with acceptance down to 5~mrad above and below the beam plane. In this configuration, one trades away the vertex capability for far better mass resolution, and performs only a resonance search. 
At the same \num{1}\% per-strip occupancy limit as the HPS SVT, a beam current of \SI{10}{\uA} on a \num{2.5}\%$X_0$ target may be used, where conservative operation at \SI{1}{\uA} allows for \SI{13}{C} on target in \SI{150}{days}, and is used to estimate the reach of this concept. While full simulation would be required to refine the background model, validate the estimated momentum and mass resolutions, and establish the techniques required for the extremely high-statistics resonance search, the right panel of Figure~\ref{fig:SuperHPS} gives an indication of the potential for a compact high-rate, high acceptance, two-armed spectrometer operating in an electron beam to probe difficult to reach parameter space for low-mass dark photons.